\documentstyle[psfig]{mn}

\title[Dwarf Nova Oscillations and Quasi-Periodic Oscillations in Cataclysmic Variables: I.]
{Dwarf Nova Oscillations and Quasi-Periodic Oscillations in Cataclysmic Variables: I. Observations of VW Hyi}
\author[Patrick A. Woudt and Brian Warner]
       {Patrick A. Woudt\thanks{email: pwoudt@artemisia.ast.uct.ac.za} and Brian Warner\thanks{email: warner@physci.uct.ac.za}\\
        Department of Astronomy, University of Cape Town, Private Bag,
        Rondebosch 7700, South Africa}
\date{}

\begin{document}

\maketitle

\begin{abstract}
From archived and recent high speed photometry of VW Hyi we find Dwarf Nova
Oscillations (DNOs) occasionally present throughout outburst, evolving from 14.06 s
period at maximum to $>$ 40 s near the end of outburst. A relatively slow increase
of period is followed by rapid increase and subsequent decrease. 

Quasi-periodic Oscillations (QPOs) are seen at periods of hundreds of seconds. For the
first time, an evolution of QPO period is seen, steadily increasing during the final
decline of an outburst. The occasional presence of two DNOs, separated in
frequency by the QPO frequency, suggests reprocessing of the rotating DNO beam
by a `wall' rotating progradely in the disc at the QPO period. 

\end{abstract}


\section{Introduction}
Dwarf nova oscillations (DNOs) were first discovered in outbursts of
the dwarf novae CN Ori and Z Cam, and in the nova-like variable UX UMa
(Warner \& Robinson 1972). They have since been observed in about
15 dwarf novae and 4 nova-likes (see Table 8.2 of Warner 1995a). 
The DNOs are low amplitude, moderately coherent luminosity variations
with periods in the range of 5--40 s. In the same kinds of cataclysmic 
variable (CV) stars there are also occasionally luminosity modulations
of longer period ($\sim$50--1000 s) and poor coherence, known as 
quasi-period oscillations (QPOs), first discussed by
Patterson, Robinson \& Nather (1977).

A variety of statistical analyses and physical models have been produced
for the DNOs and QPOs (see review by Warner 1995a), but no generally 
accepted model yet exists. Although there has been significant progress,
at least in restricting possible models, by observations made in the
far ultraviolet, it is still the case that further informative observations
are to be encouraged. 
Here we present new and archival observations of VW Hyi, and discuss
their implications. But first, in Sections 2 and 3, we review the status
quo of observations and interpretations of DNOs and QPOs.  In Section 4 we
present and analyse the observations of VW Hyi. In a subsequent paper
(Warner \& Woudt 2002, hereafter Paper II), we develop a model to explain
these observations, and apply it to other systems.

\section{DNO Phenomenology}

In the optical, DNOs are low amplitude, usually sinusoidal modulations in
brightness of moderate stability (`Q' factors = $\dot{P}^{-1}$ of 10$^4$--10$^6$). Although
not present in all dwarf nova outbursts, when they are observable they usually
appear about midway up the rising branch of outburst and disappear at a 
comparable brightness on the descending branch. Their coherence is maximal at 
the brightest phase of outburst; on the descending branch they become less
coherent and difficult to detect above the noise in the Fourier transforms.

There is a very strong correlation between oscillation period $P$ and system
brightness (Warner \& Robinson 1972), such that $P$ passes through a minimum
about one day after visual maximum, which is when the EUV luminosity 
reaches maximum (Mauche 1996a,b). At this time, the rate of mass transfer $\dot{M}$ in
the inner disc (and onto the white dwarf primary) reaches its maximum
(Cannizzo 1993). Both the short periods of the DNOs and their correlation
with $\dot{M}$ in the inner disc show that they have their origin near the
surface of the primary (Warner 1995b).

The amplitude and phase changes of the optical DNOs observed during eclipses
show that in the optical the entire disc is in some manner involved. This is
interpreted as implying that modulated high energy radiation from the central
regions of the disc is reprocessed by the whole disc. The phase variations 
can be understood only by an anisotropic radiation pattern revolving at the
period of the DNOs (Warner 1987); this limits physical models to bright regions
revolving on the surface of the primary or in the inner disc.

Optical DNOs in some nova-like variables are intermittently present (e.g.~in
UX UMa, HL Aqr, V3885 Sgr), with periods that wander slowly over ranges of a
few seconds (Warner \& Nather 1972; Knigge et al.~1998). DNOs are, however,
not ubiquitous: despite intensive observation none have been detected in
RW Sex, nor in some other less well-observed nova-likes.

DNOs have been observed in the soft X-ray region of SS Cyg, VW Hyi, SU UMa
and HT Cas during outbursts (see Warner 1995a for references; and van Teeseling 1997),
in the EUV of SS Cyg (Mauche 1996a,b, 1998) and in the UV of UX UMa
and OY Car with the use of HST (Knigge et al.~1998; Marsh \& Horne 1998).
In general the oscillations are monoperiodic and sinusoidal; recently noted
exceptions are OY Car where two periodicities are observed, one of which 
has a strong harmonic (Marsh \& Horne 1998), SS Cyg where a low amplitude
first harmonic is seen (Mauche 1997) and V2051 Oph where two periods and a 
harmonic are seen (Steeghs et al.~2001). At times, the soft X-ray modulations can
be as large as 100\%.

The temporal behaviour of DNOs, as observed both in optical and X-rays, has
features which constrict possible models. In dwarf nova outbursts, superposed
on the luminosity-related systematic changes in $P$ are intervals of 1--2 h 
during which a constant period (with some phase noise) exists, followed by an
abrupt change to another stable period. This behaviour is well illustrated
in Fig.~13 of Warner, O'Donoghue \& Wargau (1989), Fig.~6 of Cordova
et al.~(1980) and Fig.~11 of Jones \& Watson (1992). As pointed out
by Warner (1995b), these abrupt changes (occurring within 100 s) are not
accompanied by any noticeable increase or decrease of luminosity and therefore cannot
be ascribed to a change in rotation period of any substantial
($\ge 10^{-12}$ M$_{\odot}$) mass. The SS Cyg X-ray observations show jumps
of up to 0.07 s (Jones \& Watson 1992); those of VW Hyi showed $P$ changing
abruptly from 14.34 to 14.26 s (van der Woerd et al.~1987); the abrupt
changes of the $\sim$25 s oscillations in TY PsA near outburst maximum are of 
$\sim$0.15 s relative to the slow secular change in $P$ (Warner et al.~1989). 
DNOs alternating between 11.58 and 11.66 s have been seen 
in RU Peg in outburst (Patterson et al.~1977).

Marsh \& Horne (1998) have found DNOs in OY Car towards the end of a 
superoutburst. Their HST observations show two periods near 18 s
simultaneously present in the UV with a separation of 0.22 s. Although a rare
occurrence, such pairs of periodicities have been seen before in optical
DNOs. We note in particular that periods of 29.08 and 30.15 s were found
in the nova-like V3885 Sgr (Hesser, Lasker \& Osmer 1974), but the former
is normally the only (if any) period present (Warner 1973); evidence
for 26.42 and 26.73 s oscillations simultaneously present in KT Per during 
outburst is given by Robinson (1973); and the dwarf nova WZ Sge in quiesence
shows 27.87 and 28.97 s periodicities, sometimes together and at other times
separately (Robinson, Nather \& Patterson 1978). Recently Steeghs et al.~(2001) found
59.54 s, 29.77 s and 28.06 s oscillations in the optical continuum of V2051 Oph
on the decline from a normal outburst. 

There is no clear
evidence for the presence of more than two periodicities at a given time: 
some early claims (Warner \& Robinson 1972) were later shown to be due to
interference effects in the periodograms of signals with systematically
changing periods (Warner \& Brickhill 1978).

Finally, we note the perplexing optical behaviour of VW Hyi near the end of
its outbursts where in one instance a 30 s DNO was observed to be modulated
in amplitude by a QPO at a period of 413 s (Warner \& Brickhill 1978) and on
another occasion a 23.6 s DNO was modulated at 253 s (Robinson \& Warner 1984).
This cannot be simply a beat phenomenon between DNO periods separated by
$\sim$2.2 s because the average (background) brightness is also modulated at
the longer QPO period. This will become clearer below when we present the 
latest observations of VW Hyi.

\section{QPO Phenomenology}

QPOs in cataclysmic variables (see the compilation in Table 8.2 of Warner 
1995a) can have a life of their own, independent of the DNOs. For example,
QPOs with periods of $\sim$75~s and $\sim$150~s have been observed in U Gem
(Robinson \& Nather 1979) for which no optical DNOs have ever been detected.
DNOs and QPOs are sometimes present at different times in the same outburst, 
e.g.~ TY PsA (Warner et al.~1989).
However, the amplitude modulation of the DNOs in VW Hyi at the QPO period
shows that some interaction between the two processes can occur. 

\begin{figure*}
\centerline{\hbox{\psfig{figure=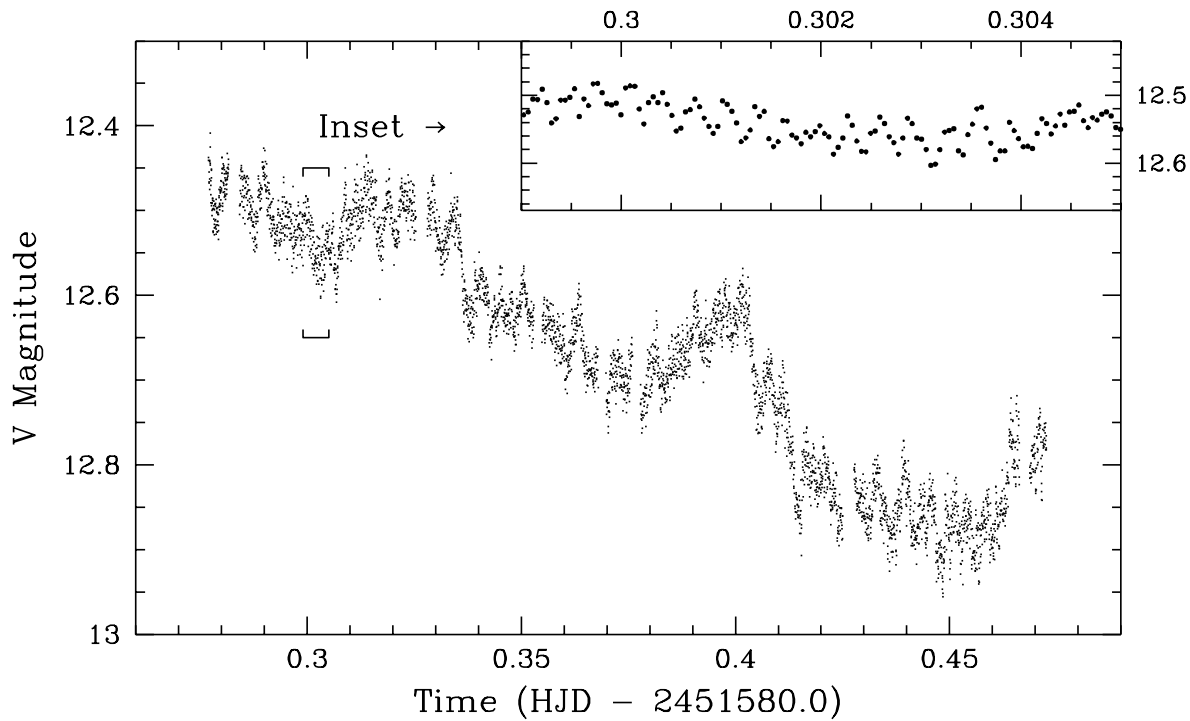,width=16cm}}}
  \caption{The light curve of VW Hyi on 5 February 2000, taken during the
late decay phase of this dwarf nova outburst. The inset is an amplified view
of a small part, showing the DNOs.}
 \label{vwfig1}
\end{figure*}

The QPO amplitudes are usually only a few $\times$ 0.01 mag, but larger
ranges have occasionally been seen. For example, the 413 s QPO in VW Hyi 
in its November 1974 outburst had a range of up to 0.12 mag (Warner \& Brickhill
1978). Similarly large QPOs with a period of 370 s have been observed in
SW UMa (Kato, Hirata \& Mineshige 1992); these comprise a nearly sinusoidal
variation of range 0.075 mag with an additional narrow dip of depth 0.05 mag.
The superoutburst of T Crv in February 2001 also showed QPOs, with period 
$\sim$600 s and range 0.1 mag (VSNET 21 Feb 2001). We show below some QPOs
in VW Hyi in which individual oscillations have a range of $\sim$0.4 mag.

The QPOs generally have average periods of 5 to 15 times the DNO periods in 
those stars where they occur simultaneously. This statement becomes more 
strongly based if it is noted that many of the shorter period ($\le 40$ s)
QPOs (Table 8.2 of Warner 1995a) should probably be reclassified as 
DNOs of low coherence.

Although QPOs are seen to change in (mean) period in a given object, unlike 
the DNOs no clear period-luminosity relationship has yet been deduced. Our
new observations of VW Hyi provide the first evidence for such a relationship.

QPOs are rarely visible at high energies: four distinct possibilities
are of 12\% amplitude at 585 s in U Gem (Cordova \& Mason 1984), of
very low amplitude at 83 s in SS Cyg (Mauche 1997), of 2240 s in OY Car
just after an outburst (Ramsay et al.~2001) and 
a modulation increasing in period from 63 s to 68 s and in amplitude from
14\% to 21\% in VW Hyi (van der Woerd et al.~1987). Wheatley et al.~(1996)
observed flux variations of large amplitude and time scale $\sim$500 s in 
Ginga (2-10 keV) X-ray observations of VW Hyi made at the end of outburst.
As this is the place in the light curve where we see optical QPOs in VW Hyi,
it is extremely probable that the X-ray observations are a different
manifestation of the same phenomenon. We return to this later.
Other possible X-ray QPOs are 290 s in AB Dra on the rise to outburst, 121
and 135 s in U Gem at quiescence and 254 s in the nova-like RW Sex (Cordova
\& Mason 1984).

It is possible that there is more than one cause for the QPOs. For example,
it has been noted that the longest QPOs have periods close to the expected 
rotation periods at the outer edges of the accretion discs in CVs (Warner 
1995a), and Lasota, Kuulkers \& Charles (1999) have suggested a model for one
of the DNOs in WZ Sge which uses a plasma blob at the disc rim.
Nevertheless, it is also possible that all QPOs are caused by 
oscillations in the inner accretion disc. Perturbation analyses of such discs
by Carroll et al.~(1985), Lubow \& Pringle (1993), and Collins, Helfer \&
van Horn (2000) show the possibility of a wide spectrum of non-radial
oscillations, analogous to p-mode and g-mode oscillations in stars. The
brightness oscillations are usually ascribed to intrinsic luminosity 
variations in the disc itself, but we point out in Paper II that they may also
be caused by variation of {\it intercepted} radiation from the central high
luminosity region of the disc, if there are vertical oscillations in
thickness of some disc annuli.

\section{Photometric Observations of VW Hyi}

The dwarf nova VW Hyi is a rich source of short period
luminosity modulations. It is the CV in which DNOs were first directly 
observable in the light curve (Warner \& Harwood 1973) (rather than only
in the Fourier transform). Unlike most dwarf novae, the optical DNOs in VW Hyi appear
most conspicuously in the final stage of decline from an outburst (they could be present
at a similar brightness at the start of an outburst, but no high time 
resolution photometry has been made that early) and follow a period-luminosity
correlation over the range 20--36 s (Warner \& Brickhill 1974, 1978; Robinson
\& Warner 1984; Haefner, Schoembs \& Vogt 1979; Schoembs \& Vogt 1980).

\subsection{Evolution of DNOs and QPOs in VW Hyi}

At the end of the November 1974 normal outburst and of the January 1978 
normal outburst the QPO modulated DNOs already mentioned above were
observed (Warner \& Brickhill 1978, hereafter WB; Robinson \& Warner 1984, hereafter RW). 
This behaviour has remained unique to VW Hyi, and is clearly of potential value in 
understanding both QPOs and DNOs. We now report a third and even more 
significant observation of this phenomenon, which has led us to reanalyse other observations
made over the past three decades.

VW Hyi was observed on 5 February 2000 with the University of Cape Town
CCD photometer attached to the 40 inch reflector at the Sutherland site
of the South African Astronomical Observatory. 3 s integrations in white light
were used, with a run length of 5h 16m, starting at 18h 40m UT. A low
time resolution light curve, corrected for atmospheric extinction, is shown
in Fig.~\ref{vwfig1}. VW Hyi was in the final stages of decline from a normal outburst
which had commenced on 3 February 2000. It was quickly apparent that in the light 
curve VW Hyi had DNOs with a period of 28~s (see the inset of Fig.~\ref{vwfig1}) 
and that these were partly modulated in amplitude with a period near 450~s, 
thus producing only the third clear example of such behaviour to be captured 
in nearly thirty years of sporadic photometry.

The general form of the light curve is shown in Fig.~\ref{vwfig1} and is typical of the 
final decay phase of a dwarf nova outburst. There was a fall of 0.45 mag during
the run; the prominent large humps are the orbital modulation (arising from 
differing aspects of a bright spot) beginning to be seen against the fading
background of the disc. A light curve of one orbit duration obtained on the following
night has a hump of approximately the same amplitude (on an intensity scale) 
as in Fig.~1, showing that there were no large changes in mass transfer rate
from the secondary in the late stages of this normal outburst. Eliminating
the orbital humps, VW Hyi was fading at 0.09 mag h$^{-1}$ on our photometric
system.

Also prominent in Fig.~\ref{vwfig1} are modulations with a time scale $\sim$7 min which are
present throughout, riding happily over the orbital humps and maintaining a peak-to-peak
range of $\sim$0.10 mag. These represent an approximately constant modulated fraction
of the declining luminosity (if Fig.~\ref{vwfig1} is plotted on an intensity scale the
$\sim$7 min modulation increases in amplitude by a factor of 2 through the run).

\begin{figure}
\centerline{\hbox{\psfig{figure=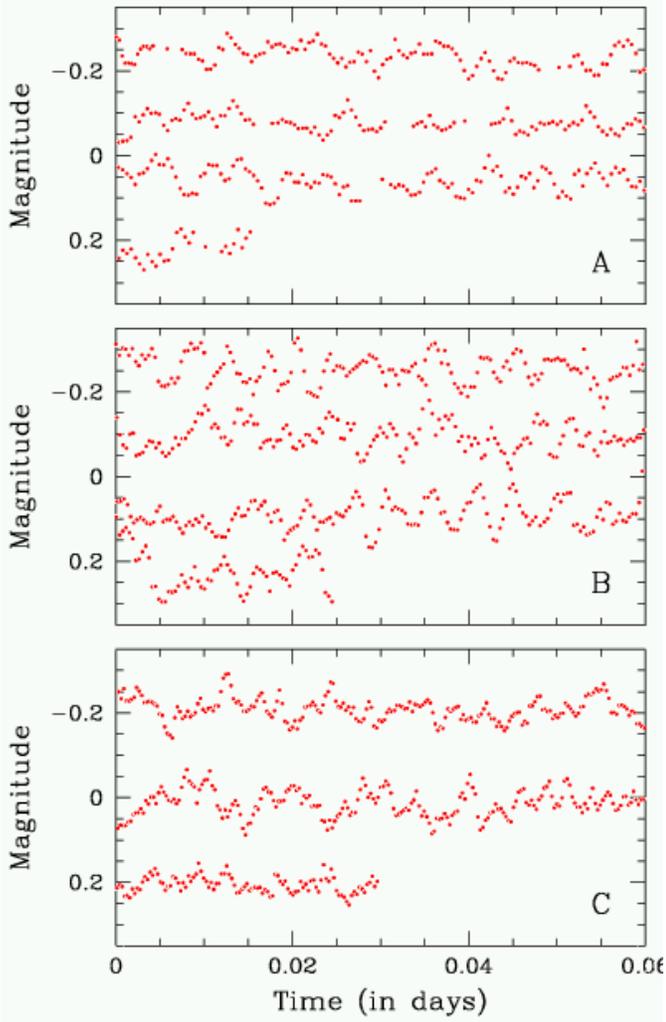,width=8.8cm}}}
  \caption{QPOs in the outbursts of (A) February 2000, (B) November 1974 and
(C) January 1978.}
 \label{qpos}
\end{figure}

\begin{figure}
\centerline{\hbox{\psfig{figure=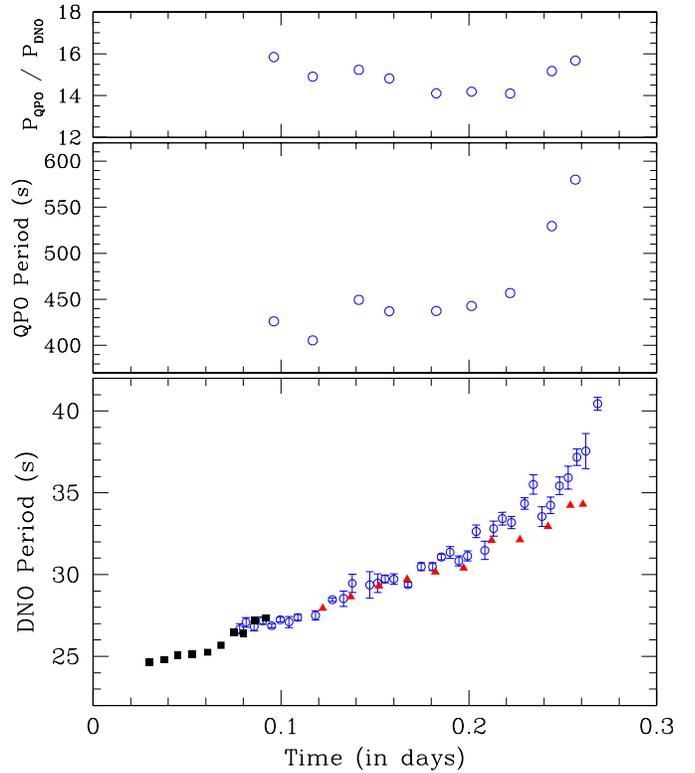,width=8.8cm}}}
  \caption{Variations with time of the DNO and QPO periods in the normal
outburst of February 2000 (circles with error bars). DNOs are added for the superoutburst of December
1972 (triangles) and the normal outburst of February 2001 (squares). 
The topmost panel shows the ratio of the periods in the February 2000 run.}
 \label{dnoevo}
\end{figure}

\begin{figure}
\centerline{\hbox{\psfig{figure=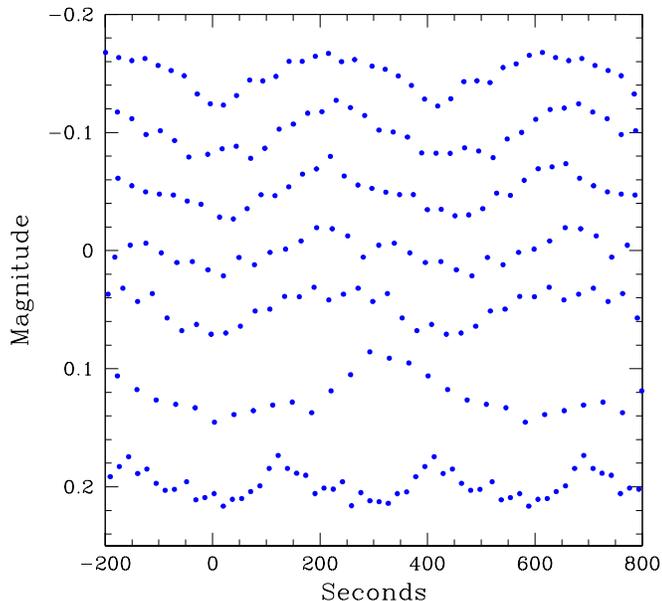,width=8.8cm}}}
  \caption{Evolution of the mean profile of QPOs in the 5 February 2000 run. Time runs from top
to bottom. Approximately five cycles are averaged and two cycles of the mean profile 
are plotted. Arbitrary vertical shifts have been applied for display purposes.}
 \label{avqpo2000}
\end{figure}

To illustrate more clearly the $\sim$7 min modulation, which is an example
of a QPO, in Fig.~\ref{qpos} we have removed the mean, first order trend and orbital
hump (represented by a sine wave and first harmonic at the orbital period) from the light curve
and have plotted mean brightnesses averaged over 10 integration bins (i.e.~30 s), 
which concomitantly smooths over the $\sim$30 s DNOs described below. 
Fourier analysis of contiguous
sections of the light curve shows that the mean period of the QPOs increases
from $\sim$400 s to $\sim$600 s during the run. This variation is displayed in
Fig.~\ref{dnoevo}, and is the first clear example of a systematic dependence
of QPO period on luminosity. However, the QPO period of the final section is omitted
because it has halved. The average pulse shapes are shown in Fig.~\ref{avqpo2000}.
In the penultimate section some first harmonic is visible; in the final section it has become
almost entirely first harmonic.

In two of the three previously published studies of VW Hyi
at the late phase of its outbursts (the November 1974 outburst: WB, and the January 
1978 outburst: RW) the
mean brightness (interpolated under the orbital humps) was nearly constant
and the QPO periods averaged 413 s and 253 s respectively with no detectable systematic
variations during the runs (which had durations of 4.91 h and 3.60 h respectively).
The background-subtracted and binned light curves, showing the QPOs, are incorporated 
into Fig.~\ref{qpos}. The mean amplitudes (half peak-to-peak) of the QPOs in those runs were
$\sim$0.03 mag and $\sim$0.02 mag respectively. In our observations of the 
February 2000 outburst the mean amplitude is $\sim$0.02 mag. In all three cases
individual cycles can have amplitudes from nearly zero to two or three times the 
average.

A third study (Warner \& Brickhill 1974) was of VW Hyi at the end of a superoutburst
and showed DNOs rapidly increasing in period. We have reanalysed the run and have 
incorporated the DNO evolution in Fig.~\ref{dnoevo}. We also find low amplitude
QPOs, not recognised in the earlier study. After the superoutburst
the DNOs lengthen in period more slowly than after the normal outburst. The ratio of
periods $P_{QPO}/P_{DNO}$ is nearly constant at $\sim 15$ in both runs.

\subsection{DNOs and QPOs in VW Hyi: the overall picture}

\begin{figure}
\centerline{\hbox{\psfig{figure=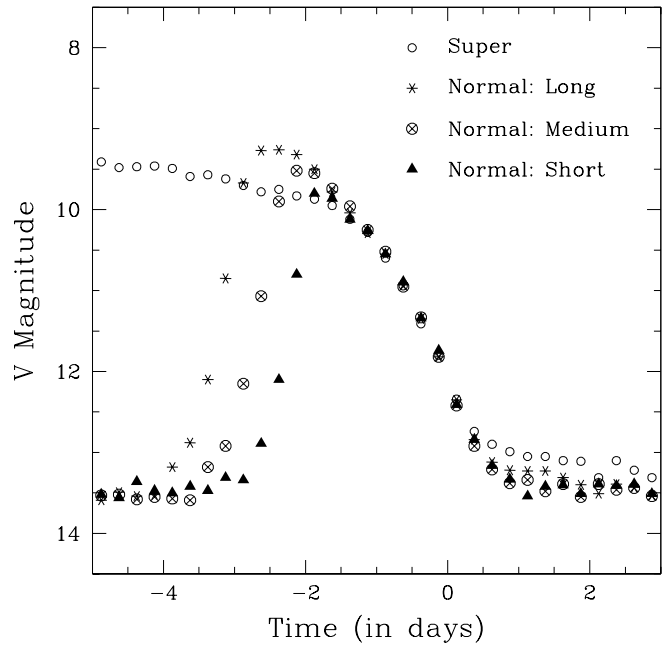,width=8.8cm}}}
  \caption{Average profiles of superoutbursts and the three types of
normal outburst.}
 \label{temp}
\end{figure}

Stimulated by the newly recognised evolution of the DNOs and QPOs we decided to
reinvestigate our archive of VW Hyi observations and to make new observations. 
In order to place each run in its respective position in the outburst we constructed a grand mean light curve.
It has long been known that the decay parts of normal outbursts of a dwarf nova are
very similar to each other (e.g.~Bailey 1975), and alignments of superoutbursts of 
VW Hyi according to the final rise to supermaximum result in nearly identical outburst
profiles (Marino \& Walker 1979). Because we are interested here largely in the decay
portion only, we have aligned these using a data file for the years 1969 -- 2000 kindly
made available by the Royal Astronomical Society of New Zealand. In performing this
analysis we found that normal and superoutburst declines closely follow the same profile.
We also found, incidentally, that the normal outbursts can be classified into three discrete
groups, of different durations. Smak (1985) found correlations of outburst widths with
outburst intervals, but not that outburst widths show signs of preference for certain
values.

Fig.~\ref{temp} shows the final `template' which we used to define where our individual 
runs are positioned. The principal portion of the decay is identical independent of the
type of outburst. The profiles near maximum, and the final decays and brightnesses at
quiescence, show small differences. The zero of the time scale is an arbitrarily chosen 
point.

\begin{table*}
 \centering
  \caption{An overview of our data archive of VW Hyi. Observations during outburst.}
  \begin{tabular}{@{}lrlrrccl@{}}
   Run        & Date & Outburst & Phase  & Length  & DNO & QPO & Remarks \\
              &      & Type     & (days) & (hours) &     &     &     \\[10pt]
 S0110 & 10 Dec 1972 & Super      &--14.85 &  2.47 &  No &  No & Early rise to supermaximum, no DNOs $>$ $1.0 \times 10^{-3}$.\\
 S0111 & 10 Dec 1972 & Super      &--14.77 &  0.97 &  No &  No & Rise to supermaximum, no DNOs $>$ $9 \times 10^{-4}$.\\
 S0112 & 10 Dec 1972 & Super      &--14.67 &  2.31 &  No &  No & Final rise to supermaximum, no DNOs $>$ $7 \times 10^{-4}$.\\
 S2230 &  9 Dec 1975 & Super      &--13.33 &  0.68 &  No & --- & No DNOs $>$ $1.0 \times 10^{-3}$.\\
 S2233 & 10 Dec 1975 & Super      &--12.40 &  0.59 &  No & --- & No DNOs $>$ $2.5 \times 10^{-3}$.\\
 S3434 & 24 Oct 1984 & Super      &--11.70 &  3 65 &  No &  No & At supermaximum, no DNOs $>$ $7 \times 10^{-4}$.\\
 S2241 & 11 Dec 1975 & Super      &--11.34 &  2.63 &  No & Yes?& No DNOs $>$ $1.8 \times 10^{-3}$, some evidence for QPO \\
       &             &            &        &       &     &     & at 745 s towards the end of the run.\\
 S3435 & 25 Oct 1984 & Super      &--10.70 &  3.68 &  No & --- & No DNOs $>$ $1.1 \times 10^{-3}$.\\
 S2243 & 12 Dec 1975 & Super      &--10.36 &  1.65 &  No & Yes & Probable QPO at 425 s. No DNOs $>$ $1.1 \times 10^{-3}$.\\
 S3436 & 26 Oct 1984 & Super      & --9.79 &  2.06 & Yes &  No & DNO at 14.29 s ($3.5 \times 10^{-3}$) in first 50 min.\\
 S0115 & 18 Dec 1972 & Super      & --6.86 &  1.20 &  No &  No & No DNOs $>$ $6 \times 10^{-4}$.\\
 S3078 & 19 Dec 1982 & Super      & --5.32 &  1.95 & Yes &  No & Stable DNO at 14.06 s, ampl.~$1.2 \times 10^{-3}$.\\
 S0480 & 30 Nov 1973 & Super      & --5.19 &  1.73 &  No &  No & No DNOs $>$ $2.5 \times 10^{-3}$.\\
 S0118 & 20 Dec 1972 & Super      & --4.86 &  1.93 &  No &  No & No DNOs $>$ $9 \times 10^{-4}$.\\
 S3437 & 31 Oct 1984 & Super      & --4.62 &  2.90 &  No & --- & No DNOs $>$ $1.2 \times 10^{-3}$.\\
 S0120 & 21 Dec 1972 & Super      & --3.87 &  2.03 &  No &  No & No DNOs $>$ $8 \times 10^{-4}$.\\
 S3692 &  8 Nov 1985 & Super      & --3.81 &  0.36 &  No & --- & No DNOs $>$ $1.4 \times 10^{-3}$.\\
 S3693 &  8 Nov 1985 & Super      & --3.80 &  0.29 &  No & --- & No DNOs $>$ $1.3 \times 10^{-3}$.\\
 S3438 &  1 Nov 1984 & Super      & --3.75 &  1.94 &  No &  No & No DNOs $>$ $1.5 \times 10^{-3}$.\\
 S2911 & 25 Nov 1981 & Normal (L) & --2.71 &  3.14 &  No &  No & Final rise to normal maximum, no DNOs $>$ $8 \times 10^{-4}$.\\
 S2621 &  3 Jan 1978 & Normal (M) & --2.26 &  2.75 &  No &  No & Final rise to normal maximum, no DNOs $>$ $7 \times 10^{-4}$.\\
 S0122 & 23 Dec 1972 & Super      & --1.82 &  2.56 &  No & Yes & QPO at 410s in first half of the run ($3 \times 10^{-3}$).\\
       &             &            &        &       &     &     & No DNOs $>$ $5 \times 10^{-4}$.\\
 S1277 & 31 Oct 1974 & Normal (?) & --1.77 &  1.78 &  No &  No & At normal maximum, no DNOs $>$ $1.3 \times 10^{-4}$.\\
 S1571 & 20 Dec 1974 & Super      & --1.04 &  1.99 &  No & Yes?& Start of fall from supermaximum plateau. Possible\\
       &             &            &        &       &     &     & QPO at 1151 s ($4.2 \times 10^{-3}$). No DNOs $>$ $1.3 \times 10^{-3}$.\\
 S6183 & 15 Feb 2001 & Normal (L) & --0.94 &  1.68 &  No &  No & No DNOs $>$ $1.3 \times 10^{-3}$.\\
 S3703 & 11 Nov 1985 & Super      & --0.83 &  1.04 &  No &  No & No DNOs $>$ $2.6 \times 10^{-3}$.\\
 S0124 & 24 Dec 1972 & Super      & --0.79 &  1.84 &  No &  No & No DNOs $>$ $8 \times 10^{-4}$.\\
 S2914 & 27 Nov 1981 & Normal (L) & --0.73 &  0.52 &  No &  No & No DNOs $>$ $2.2 \times 10^{-3}$.\\
 S3410 & 22 Sep 1984 & Normal (M) & --0.58 &  0.51 &  No & --- & No DNOs $>$ $1.7 \times 10^{-3}$.\\
 S1307 &  2 Nov 1974 & Normal (M) & --0.3$^{\star}$ &  1.33 & Yes & Yes & QPOs at $\sim$185 s ($3.5 \times 10^{-3}$). DNOs at $\sim$18.2 s, frequent \\
       &             &            &        &       &     &     & small period changes. Average $2 \times 10^{-3}$, max.~$8 \times 10^{-3}$.\\
 S0018 & 11 Sep 1972 & Normal (L) & --0.11 &  2.04 & Yes &  No & DNOs lengthening (20.2 -- 20.6 s, ampl.~$2.2 \times 10^{-3}$).\\
 S1594 & 21 Dec 1974 & Super      & --0.05 &  2.77 &  No &  No & No DNOs $>$ $1.3 \times 10^{-3}$.\\
 S6184 & 16 Feb 2001 & Normal (L) &   0.06 &  1.68 & Yes & Yes & DNO evolution (24.6 $\rightarrow$ 27.4 s), see discussion in text.\\
 S2915 & 28 Nov 1981 & Normal (L) &   0.10 &  0.73 & Yes &  No & Average DNO at 21.3 s, short coherence ($\sim$660 s). Range \\
       &             &            &        &       &     &     & in DNO period 20.6 -- 22.4 s.\\
 S0127 & 25 Dec 1972 & Super      &   0.17 &  3.77 & Yes & Yes & DNO evolution (28 $\rightarrow$ 34 s), see discussion in text.\\
 S6059 &  5 Feb 2000 & Normal (M) &   0.18 &  5.27 & Yes & Yes & DNO (27 $\rightarrow$ 40 s) / QPO evolution, see discussion in text.\\
 S6138 & 19 Dec 2000 & Normal (M) &   0.54 &  7.63 & Yes & Yes & DNOs in range 25 -- 34 s of short coherence ($\sim$1260 s).\\
 S3416 & 23 Sep 1984 & Normal (M) &   0.56 &  1.19 & Yes & Yes & DNOs in range 25 -- 30 s of short coherence. Modulation \\
       &             &            &        &       &     &     & at QPO period of 300 s (see text).\\
 S1322 &  3 Nov 1974 & Normal (M) &   0.7$^{\star}$ &  4.91 & Yes & Yes & DNOs in range 26 -- 33 s. See WB.\\
 S5248 &  6 Nov 1990 & Normal (M) &   0.76 &  4.66 & Yes & Yes & QPO at 2100 s + first and second harmonic, see text.\\
       &             &            &        &       &     &     & Occasional DNOs near 40 s of low coherence.\\
 S2623 &  6 Jan 1978 & Normal (M) &   0.78 &  3.60 & Yes & Yes & DNOs in range 22 -- 27 s. See RW.\\
 S0484 &  6 Dec 1973 & Super      &   0.79 &  3.91 & Yes & Yes & Strong QPO at 1326 s ($4.4 \times 10^{-2}$), see Fig.~\ref{avQPO}.\\
       &             &            &        &       &     &     & DNOs in range of 29 -- 38 s of short coherence, see text.\\
 S0019 & 12 Sep 1972 & Normal (L) &   0.94 &  4.30 & Yes & Yes & Large amplitude QPOs at $\sim 500$ s ($2.5 \times 10^{-2}$). \\
 S1616 & 22 Dec 1974 & Super      &   0.96 &  2.27 &  No &  No & No DNOs $>$ $2.5 \times 10^{-3}$.\\
 S0129 &  8 Jan 1973 & Normal (S) &   1.04 &  2.14 &  No &  No & No DNOs $>$ $4.5 \times 10^{-3}$.\\
 S6060 &  6 Feb 2000 & Normal (M) &   1.11 &  1.53 &  No &  No & No DNOs $>$ $5 \times 10^{-3}$.\\
 S0026 & 13 Sep 1972 & Normal (L) &   1.92 &  3.32 &  No & Yes & QPOs at 1043 s ($2.4 \times 10^{-2}$). \\
 S0128 & 27 Dec 1972 & Super      &   2.14 &  1.86 &  No &  No & No DNOs $>$ $4 \times 10^{-3}$.\\
 S3715 & 14 Nov 1985 & Super      &   2.26 &  1.27 & Yes & Yes & DNOs at 24.7 s ($3.4 \times 10^{-3}$) and QPOs at $\sim$360 s \\
       &             &            &        &       &     &     & in last hour of run.\\
 S2917 & 30 Nov 1981 & Normal (L) &   2.34 &  2.18 &  No &  No & No DNOs $>$ $2.6 \times 10^{-3}$.\\
 S0030 & 14 Sep 1972 & Normal (L) &   2.96 &  6.92 &  No & Yes & Evidence for QPO behaviour, but of low coherence.\\
       &             &            &        &       &     &     & Several cycles at 1140 s ($2.5 \times 10^{-2}$).\\
\end{tabular}
\label{tab1}
\end{table*}
\addtocounter{table}{-1}

\begin{table*}
 \centering
  \caption{Continued: Selected observations during quiescence.}
  \begin{tabular}{@{}lrlrrccl@{}}
   Run        & Date & Outburst & Phase  & Length  & DNO & QPO & Remarks \\
              &      & Type     & (days) & (hours) &     &     &     \\[10pt]
 S0077 & 11 Oct 1972 & Quiescence &         &  4.11 &  No & Yes?& Some evidence for occasional QPO at 935 s.  \\
       &             &            &         &       &     &     & No DNOs $>$ $4 \times 10^{-3}$.\\
 S0085 & 13 Oct 1972 & Quiescence &         &  1.67 &  No & Yes?& Evidence for few cycles of QPO at 720 s. \\
       &             &            &         &       &     &     & No DNOs $>$ $6 \times 10^{-3}$.\\
 S0093 & 14 Oct 1972 & Quiescence &         &  3.33 &  No &   ? & No DNOs $>$ $3.5 \times 10^{-3}$. \\
 S0073 & 26 Nov 1972 & Quiescence &         &  2.83 &  No & Yes?& Possible QPO at 260 s. No DNOs $>$ $5 \times 10^{-3}$.\\
 S0102 &  5 Dec 1972 & Quiescence &         &  1.31 &  No & Yes & QPO at 833 s + first harmonic. No DNOs $>$ $4.5 \times 10^{-3}$.\\
 S0105 &  8 Dec 1972 & Quiescence &         &  1.80 &  No & Yes & QPO near 600 s + first harmonic ($2.5 \times 10^{-3}$). \\
       &             &            &         &       &     &     & No DNOs $>$ $2.5 \times 10^{-3}$.\\
 S1414 &  2 Dec 1974 & Quiescence &         &  2.89 &  No & Yes?& Evidence for QPO of short coherence at 980 s.\\
       &             &            &         &       &     &     & No DNOs $>$ $2.5 \times 10^{-3}$.\\
\hline
\end{tabular}
{\hskip 4.8cm \footnotesize $\star$ Our relative magnitudes taken from photoelectric photometry have been used to determine the 
phase of our observations on these dates with respect to the outburst template. This outburst was sparsely sampled by the observers who reported
to the RASNZ.}
\label{tab1c}
\end{table*}

\subsubsection{The DNOs}

In the process of analysing the runs we discovered previously overlooked features.
Table~\ref{tab1} lists the runs which had significant results -- including absence of 
DNOs (our upper limit on the DNO amplitude corresponds approximately to 3$\sigma$ 
significance) during outburst. We have omitted many of the runs made during
quiescence which showed neither DNOs nor QPOs.

\begin{figure}
\centerline{\hbox{\psfig{figure=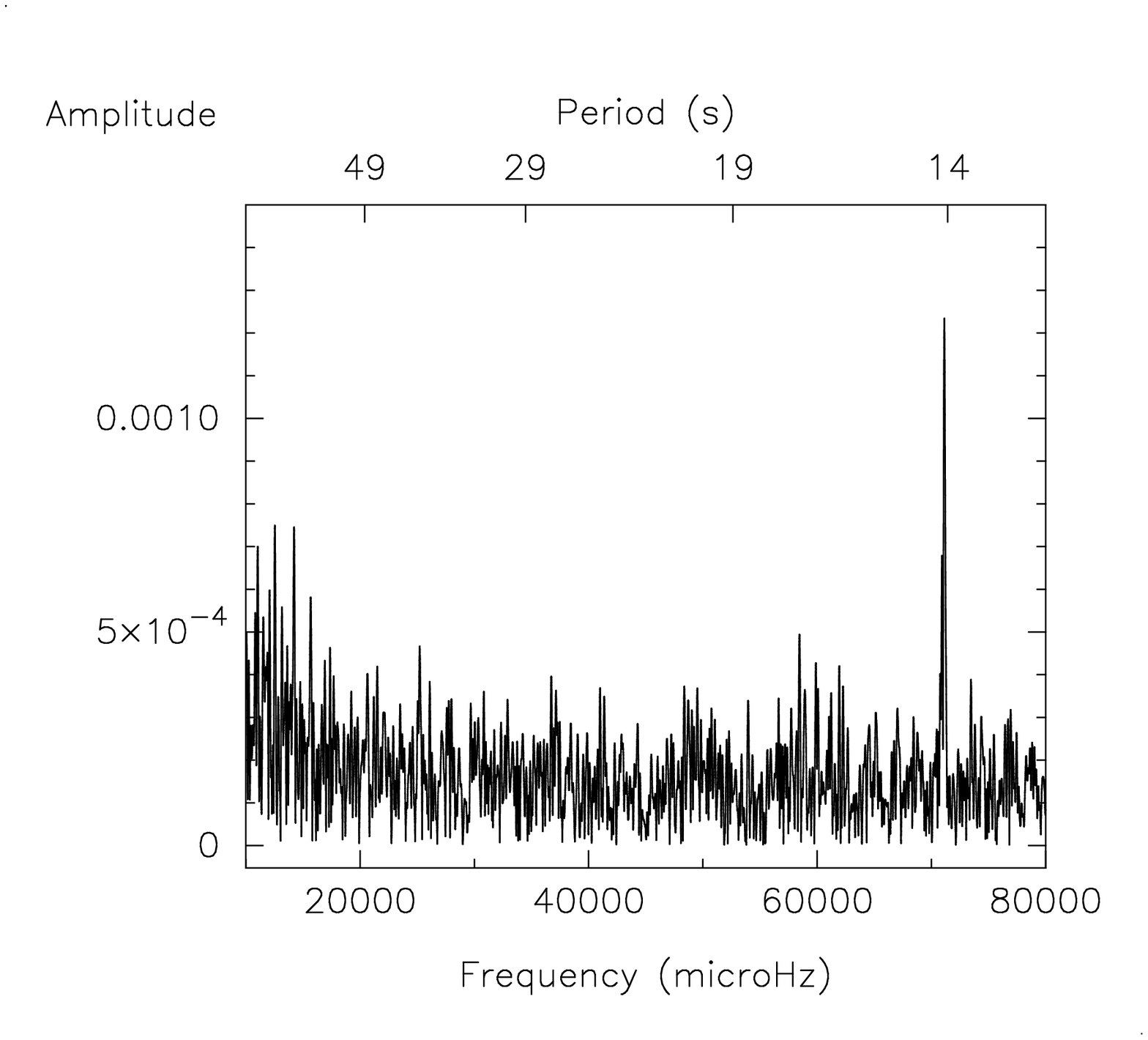,width=8.8cm}}}
  \caption{Fourier amplitude spectrum of the light curve of VW Hyi on 19 December 1982.}
 \label{s3078ps}
\end{figure}

\begin{figure}
\centerline{\hbox{\psfig{figure=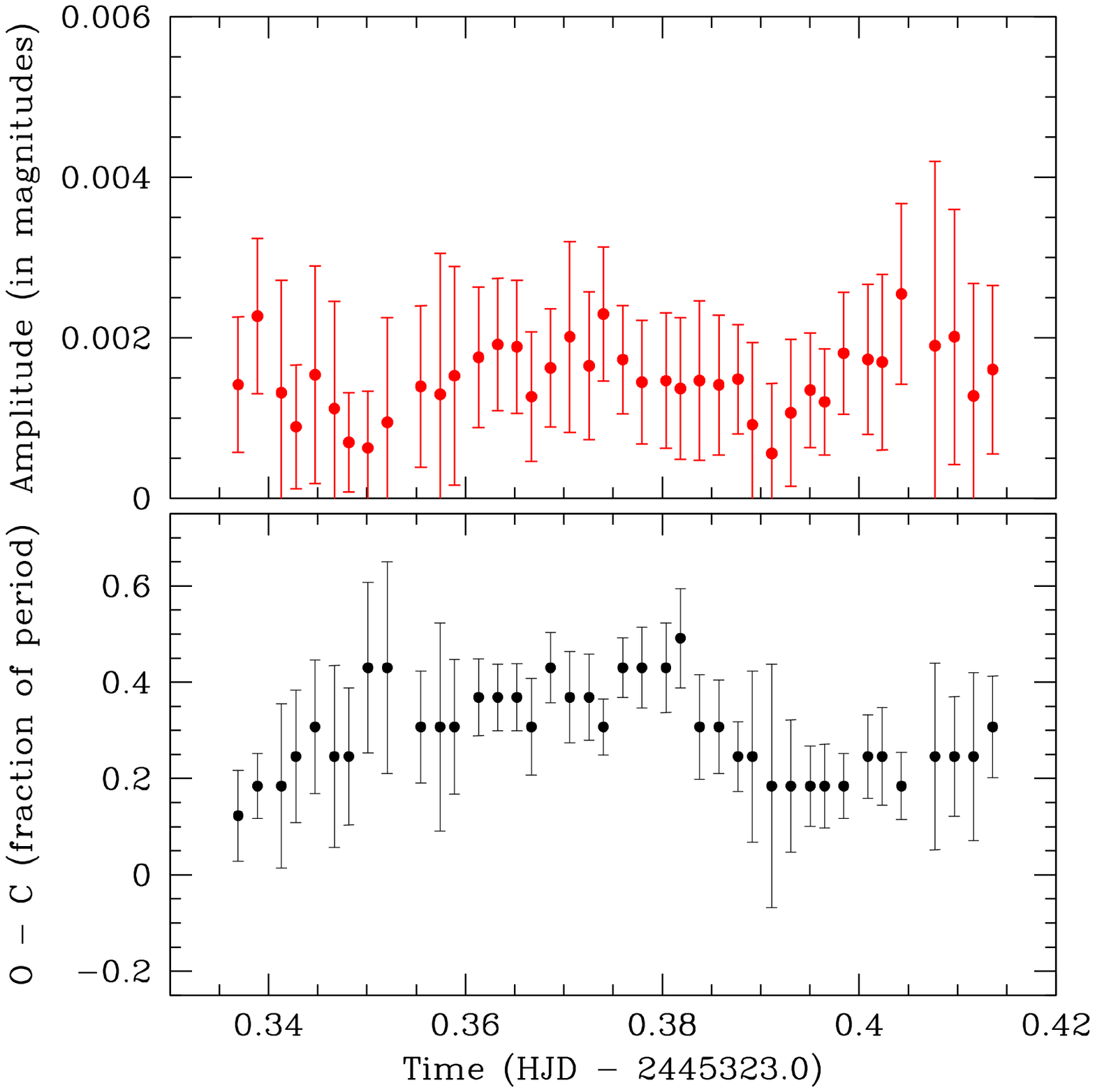,width=8.8cm}}}
  \caption{O-C diagram and amplitude variations of the 14.06 s period in the light curve 
of VW Hyi on 19 December 1982.}
 \label{omc3078}
\end{figure}

The first notable result is the presence of a DNO with period 14.06 s in the run
made on 19 December 1982, on the plateau of a superoutburst at V $\simeq$ 9.5, 9 days
after the rise to maximum. Interestingly, it is at precisely the same period as that seen
in the soft X-rays at V = 9.4 in the November 1983 superoutburst of VW Hyi (van der Woerd et al.~1987),
which will be discussed in Section 3 of Paper II. Schoembs \& Vogt (1980) found a 33.9 s DNO in the superoutburst
of VW Hyi of 27 October 1978, also at V $\simeq$ 9.5. This perplexed us at first, until we noticed
that their time resolution was 10 s, which would place a 14 s period above the Nyquist frequency -- but 
the beat period between 10 s and 33.9 s is 14.18 s, so there evidently was a 14.18 s DNO 
which appeared as a beat at considerably reduced amplitude. We also found a DNO at 14.29 s present for a 
short time in the superoutburst of 26 October 1984. The optical and X-ray coverage of VW Hyi has
been quite extensive, so we conclude that these very stable 14 s DNOs are rare and of short duration.

Our Fourier transform for 19 December 1982 is shown in Fig.~\ref{s3078ps}, in which the spike at 14.06 s is 
very narrow, denoting a stable period over the $\sim$2 h run. There is no significant power at the
subharmonic or the first harmonic. In Fig.~\ref{omc3078} we show O--C and amplitude variations, 
relative to a constant period of 14.06 s. Although there are slow variations in phase, these
are limited in range and show none of the sudden jumps in period discussed below. This is
the most stable DNO that we have observed in VW Hyi.
The amplitude ($\sim$0.001 mag) is very low and would not be detected if the
signal were much less stable. It is possible, therefore, that $\sim$14 s
modulation is commonly present at maximum light but not detectable with
common Fourier techniques.

There is a suggestion of sinusoidal variation of O--C in Fig.~\ref{omc3078} with a period
near the orbital period. The phasing relative to the position of the white dwarf in its orbit
is in rough agreement with what might be expected of a light travel time effect, but the 
expected O--C range of $2 a q {\sin} i / c (1+q)$, where $q \simeq 0.19$ is the mass ratio,
is only $\sim$0.032 P, whereas we see a range of $\sim$0.2 P. 
The O--C variations must therefore be ascribed to small intrinsic variations in DNO period
over the 2 h run.

We also found three other examples of DNOs at periods (18 -- 22 s), shorter than any
hitherto recorded in optical observations.

\begin{figure}
\centerline{\hbox{\psfig{figure=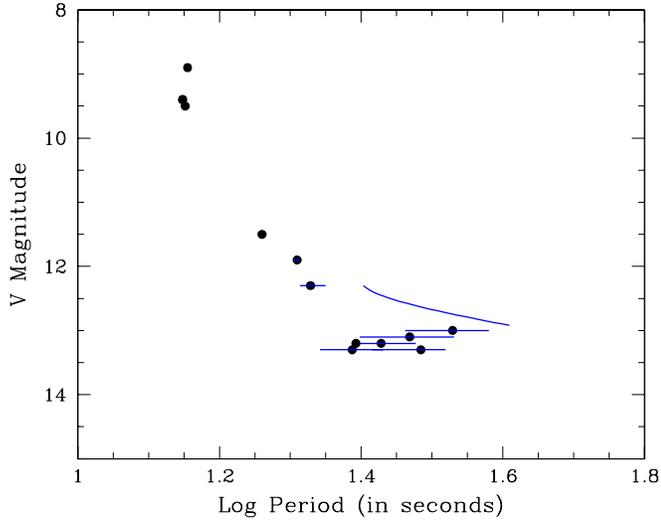,width=8.8cm}}}
  \caption{DNO periods as a function of the V magnitude of VW Hyi. The curved continuous line
corresponds to the DNO evolution seen in Fig.~\ref{dnoevo}. The horizontal bars show the range
of DNO periods in each of the runs illustrated.}
 \label{maglogp}
\end{figure}

In Fig.~\ref{maglogp} we show the evolution of DNO periods through decline of outburst, based on
the data in Table~\ref{tab1} and in Section 4.1. We have included the X-ray period (van der Woerd et al.~1987)
and Schoembs \& Vogt's (1980) beat period in this diagram. For V $\le$ 12.5 there is a clear correlation
between brightness and period, similar to what is commonly seen in dwarf nova outbursts. The
slope of this relationship is $\alpha = {\rm d} {\log}P / {\rm d} {\log}L_{opt} \simeq -0.15$, which is 
just within the range $\alpha = -0.25 \pm 0.1$ seen in other systems (Warner 1995a). However, as VW Hyi fades
through V = 12.5 the rapid increase in period illustrated in Fig.~\ref{dnoevo} takes over and the
slope steepens to at least $\alpha \simeq -2$ when $P$ has increased to 40 s. We have
not yet caught the end of this short-lived phase. Subsequently,
shorter periods ($\sim$22 -- 35 s) are seen, but whether there is a frequency doubling, and whether
different outbursts behave differently, we do not have sufficient data to judge.
At this stage of outburst (the final approach to quiescence) the DNOs are very incoherent.

\begin{figure}
\centerline{\hbox{\psfig{figure=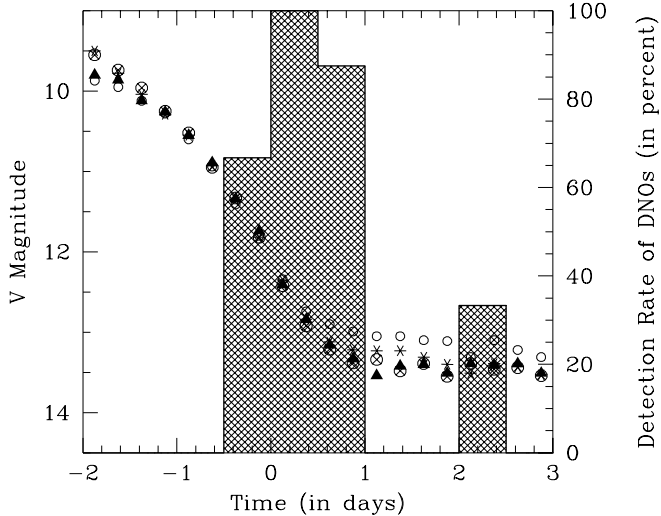,width=8.8cm}}}
  \caption{The detection rate of DNOs as a function of the relative phase in the outburst. The 
symbols used for the template light curve are as in Fig.~\ref{temp}.}
 \label{dnofreq}
\end{figure}

\begin{figure}
\centerline{\hbox{\psfig{figure=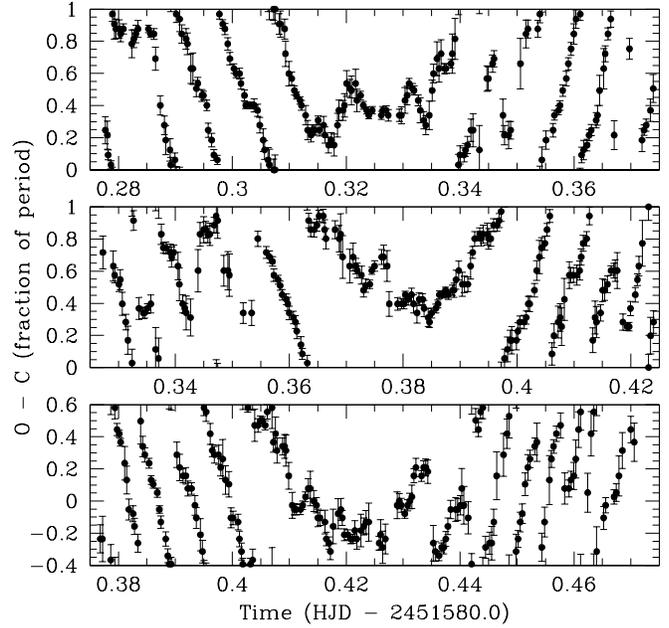,width=8.8cm}}}
  \caption{Three O--C diagrams for the DNOs present in Fig.~\ref{vwfig1}. The `oak panel' effect
is obtained by analysing separately the first half of the run (top), the central half (middle)
and the final half (bottom).}
 \label{oakpanel}
\end{figure}

Our observations provide the frequency histogram shown in Fig.~\ref{dnofreq}, which displays
the fractional success rate for detection of DNOs on the decline of the outburst. In the
range $12.5 \la {\rm V} \la 13.3$ DNOs are almost always present. Earlier and later in outbursts
their occurence falls away rapidly.

The DNOs present in the February 2000 light curve present the highest signal-to-noise and longest
data train available to us. The detailed evolution of these DNOs is shown in Fig.~\ref{oakpanel}
in which the upper and the lower panels show O--C curves, of $\sim$4 cycles of DNO with
75\% overlap, relative to constants periods of 28.0 s and 33.0 s respectively for the first and last
halves of the run. The central panel illustrates the central 50 percent, calculated with respect
to a 30.5 s period. A few very uncertain points (because of small oscillation amplitude) have been
omitted.

\begin{figure}
\centerline{\hbox{\psfig{figure=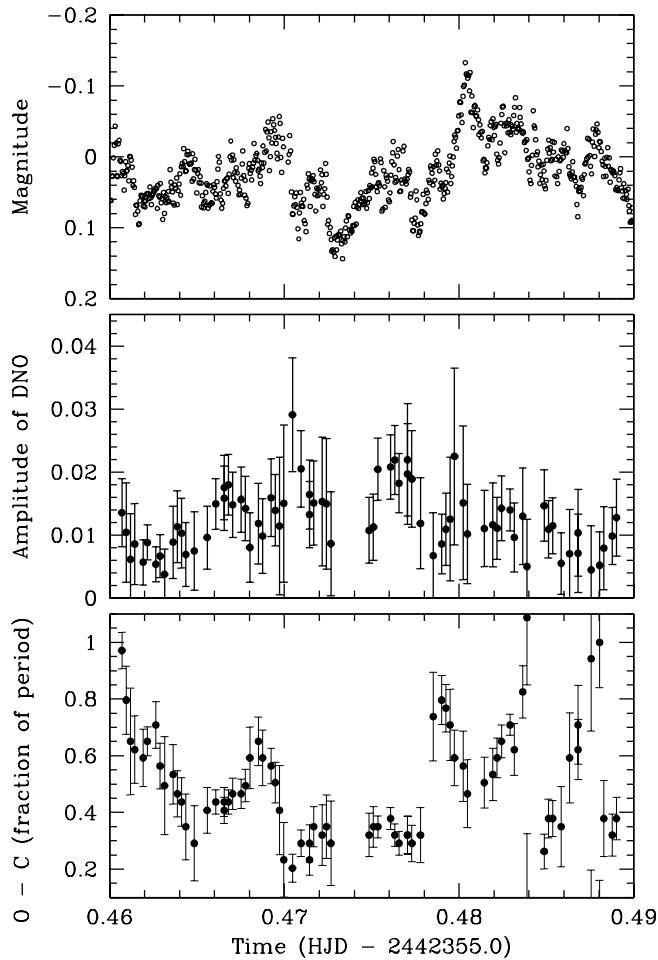,width=8.8cm}}}
  \caption{Analysis of a section of the 3 November 1974 light curve. The upper panel shows
the relative magnitudes. The central panel shows the amplitude of the DNOs obtained by fitting
a sine wave of period 29.68 s by least squares to about 5 cycles with 75\% overlap for consecutive
points. A few points of low amplitude and hence large uncertainty have been removed. The lowest panel
shows the O--C phase variations. }
 \label{phase}
\end{figure}

The quasi-cyclical variations of O--C that are readily visible in the central section
of each panel are present throughout the run, and may be detected in the compression and
rarefaction of points along the curves. These variations often correlate with QPO modulations
and are described in Section 4.2.3. There are also, in this run as in most of our DNO
observations, abrupt changes of period and occasional abrupt phase shifts. We illustrate
in Fig.~\ref{phase} some examples of these discontinuities, which are of the same kind as already
mentioned for other CVs in Section 2. There are abrupt period changes at times 0.4645 and 0.4810,
and phase discontinuities at 0.4690 and 0.4780. After 0.4810 the change in period is so large that
the O--C values run off the top of the panel and `wrap around'.
The O--C variations in this example do not correlate with the general brightness changes.

\subsubsection{The QPOs}

\begin{figure}
\centerline{\hbox{\psfig{figure=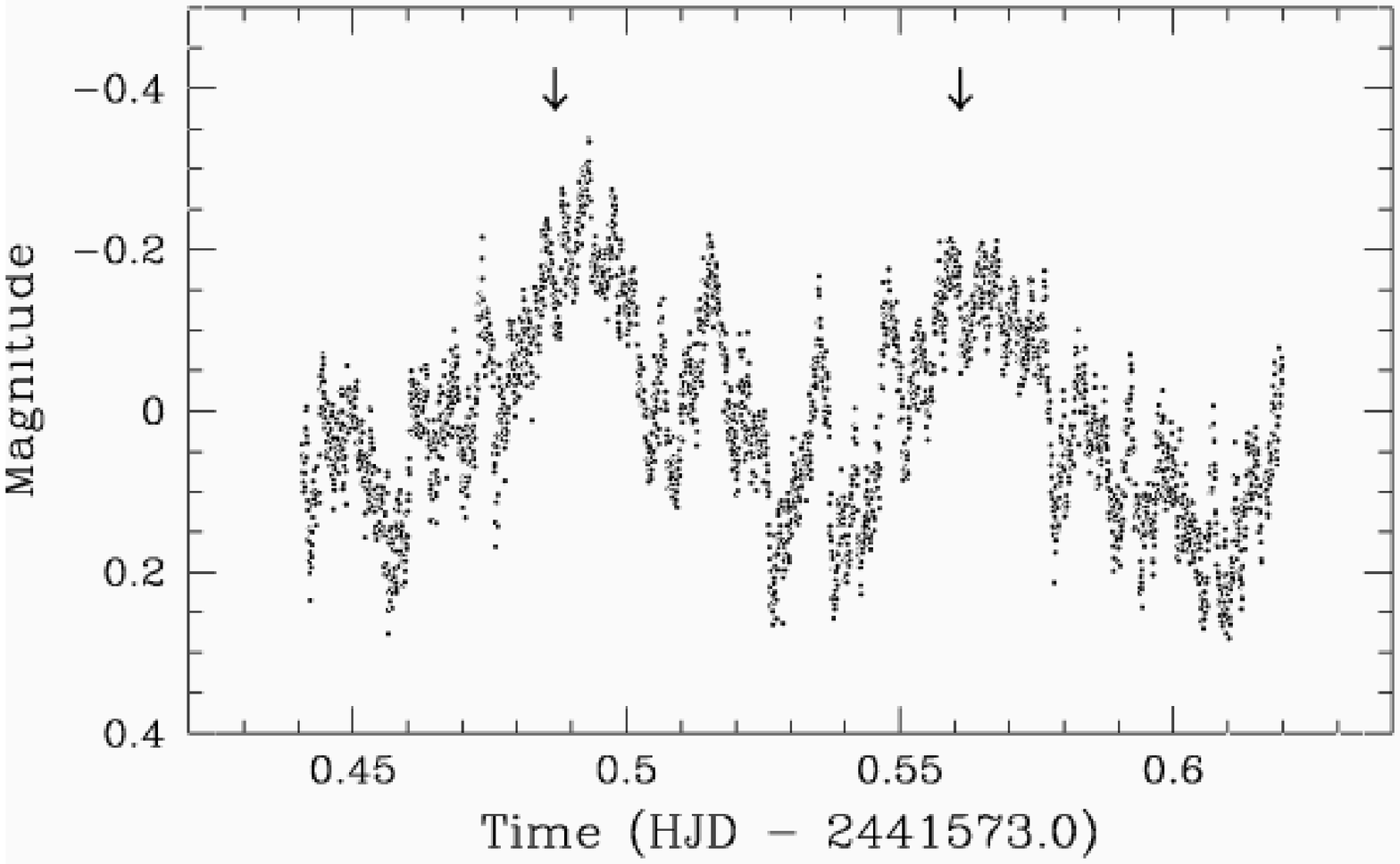,width=8.8cm}}}
\centerline{\hbox{\psfig{figure=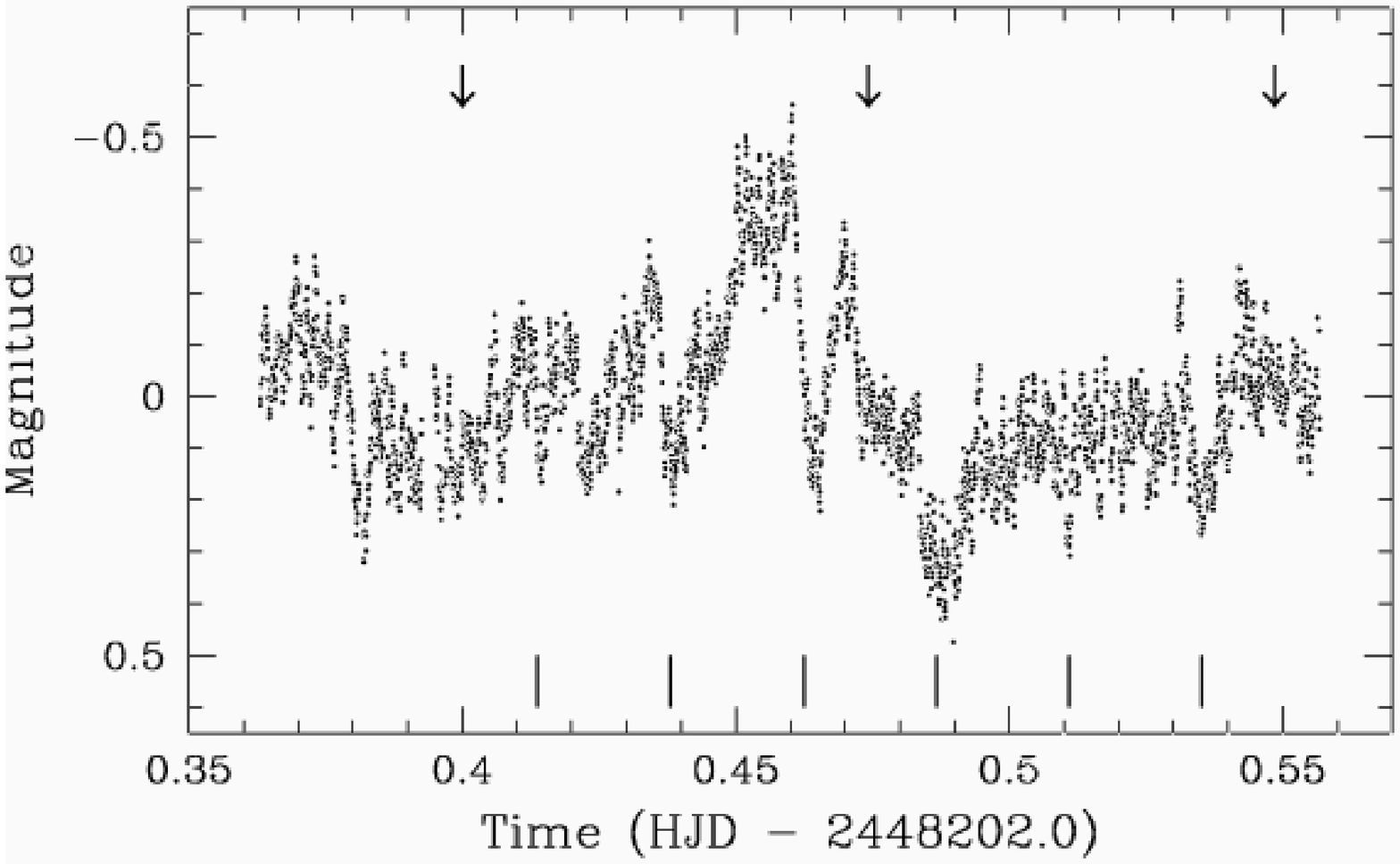,width=8.8cm}}}
  \caption{Light curve of VW Hyi on 12 September 1972 (upper panel) and
on 6 November 1990 (lower panel). Arrows indicate predicted times of orbital
hump maxima. In the lower panel, vertical bars mark the position of recurrent dips at a 
period of 2100 s.}
 \label{lc0019}
\end{figure}

\begin{figure}
\centerline{\hbox{\psfig{figure=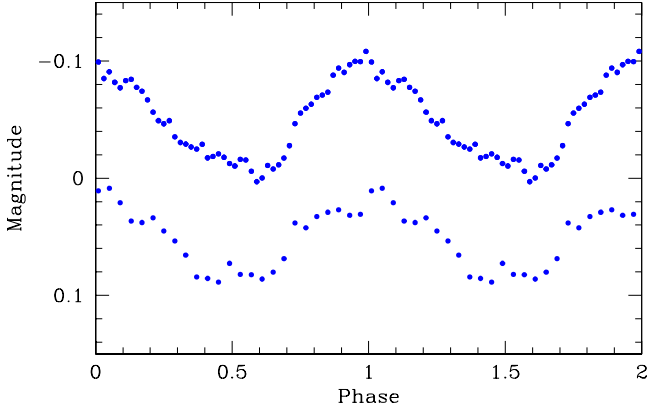,width=8.8cm}}}
  \caption{Average QPO profiles for the 6 December 1973 (upper panel) and 12 September 1972 (lower panel)
runs. The profiles are displaced by +0.05 and -0.05 mag.}
 \label{avQPO}
\end{figure}

A few runs have QPOs of large amplitude. These are listed in Table~\ref{tab1} and two
average profiles are shown in Fig.~\ref{avQPO}.
Although these QPOs are obvious in the light curves, their presence in VW Hyi (and, 
by implication, in the light curves of other CVs) has previously been ascribed to 
slow flickering. The most extreme examples are shown in the upper panel of Fig.~\ref{lc0019}, which 
shows the second run on VW Hyi made by the senior author at a time when QPOs had yet to be identified 
by Patterson et al.~(1977), and which was used merely as part of the series
of runs which first disclosed orbital modulation in VW Hyi (Warner 1975). The coherence of
the apparent large flares and dips in the upper panel of Fig.~\ref{lc0019} can be judged from the mean light curve 
(the lower profile) given in Fig.~\ref{avQPO}. 

\begin{figure}
\centerline{\hbox{\psfig{figure=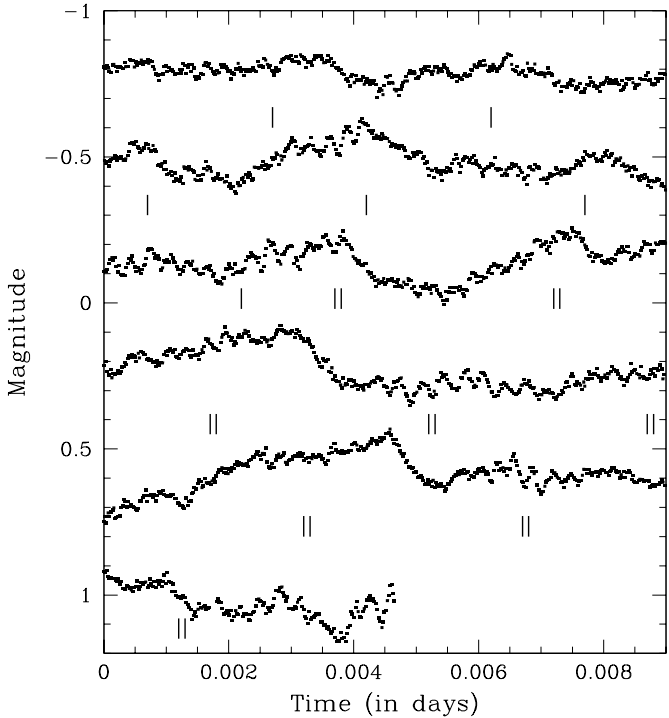,width=8.8cm}}}
  \caption{A detailed view of the DNO and QPO modulations of VW Hyi on 23 September 1984. 
The light curve spans 0.05 d and is shown in six segments displaced vertically, starting from the top. 
The QPO marked by single vertical bars, after about 0.02 d, changes phase and is then marked by double
vertical bars (see text).}
 \label{dnobup}
\end{figure}

The question of coherence is an important one. By their very nature, QPOs of short coherence
are difficult to detect in the Fourier transform. Fig.~\ref{dnobup} shows the details of the
71 min of light curve obtained on 23 September 1984 near the end of a normal outburst.
The QPO maxima, spaced 300 s apart, are shown by vertical bars. We can interpret the evolution
of the QPO in this light curve as the growth and decay of a QPO (indicated by single vertical bars)
over about 5 cycles, followed by growth and decay over 4 - 6 cycles of another QPO (double
bars) of similar period, but phase shifted relative to the first QPO by $\sim$0.4 cycle.
Such a phase shift leads to spread of power and lowering of peak amplitude in the Fourier transform.

A characteristic of large QPOs is that at their minima they drag the intensity well
below the smooth lower envelope of the light curve.

\begin{figure}
\centerline{\hbox{\psfig{figure=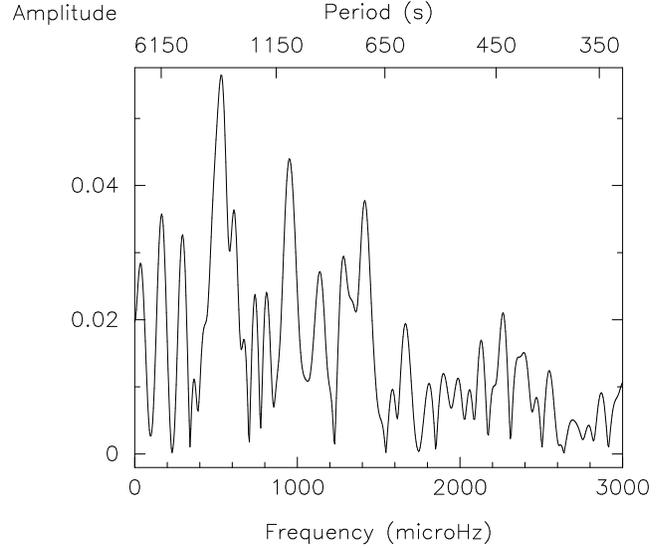,width=8.8cm}}}
  \caption{The Fourier spectrum of the prewhitened light curve of 6 November 1990. The fundamental and
the first harmonic of the orbital period have been removed.}
 \label{four5248}
\end{figure}

\begin{figure}
\centerline{\hbox{\psfig{figure=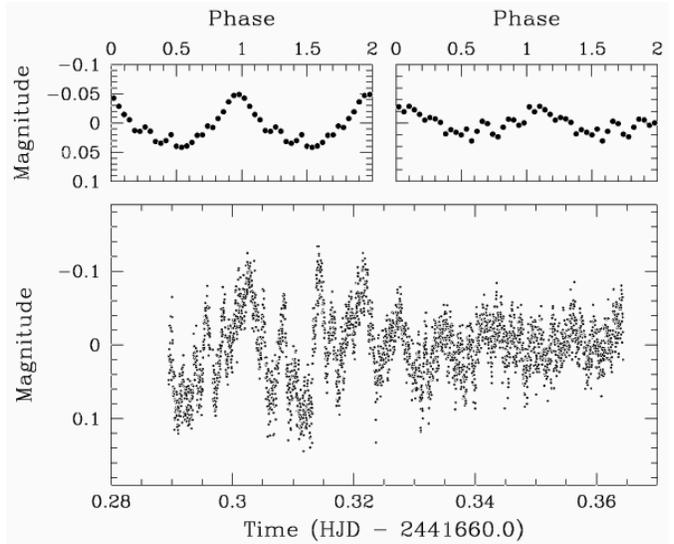,width=8.8cm}}}
  \caption{Light curve of VW Hyi in quiescence on 8 December 1972. The upper two panels show
the mean QPO profile in the first and second half of the run. }
 \label{QPOqui}
\end{figure}

\begin{figure}
\centerline{\hbox{\psfig{figure=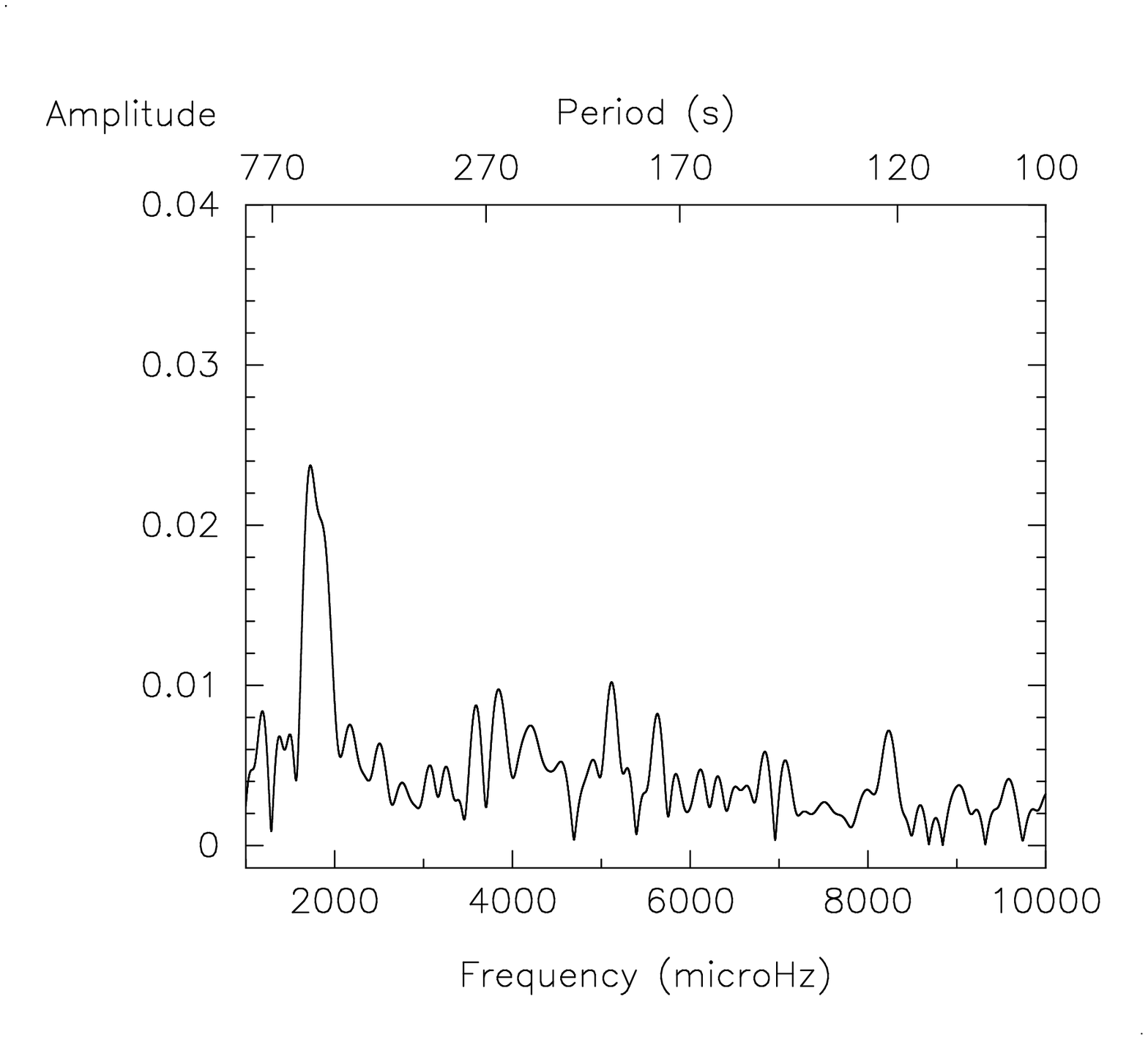,width=8.8cm}}}
  \caption{The Fourier spectrum of the light curve of VW Hyi in quiescence on 8 December 1972.}
 \label{four0105}
\end{figure}

We illustrate in the lower part of Fig.~\ref{lc0019} another light curve, obtained near the end of a
normal outburst, in which the QPO phenomenon is very strong as judged by the dips and flares. 
The Fourier transform of this light curve, Fig.~\ref{four5248}, shows the fundamental, first and second
harmonics of a period near 2100 s, which account for the repetitive narrow dips marked in Fig.~\ref{lc0019}.
The predicted times of orbital
hump maxima are shown in the light curve of 6 November 1990, the first predicted hump is absent and the third
is of low amplitude. Rapid changes in hump size were also seen by WB and imply large variations
of rate of mass transfer from the secondary at the end of outburst, perhaps the result of 
searching for stability after enduring a slightly increased rate through the effect of irradiation
during the outburst.

The final entries in Table~\ref{tab1} list seven light curves in which, in the light of the
experience of analysing QPOs in outburst, we see or suspect QPOs in VW Hyi at quiescence. This is the
first claim of the presence of QPOs in quiescent light curves of a dwarf nova. We illustrate one of 
these light curves in Fig.~\ref{QPOqui}, and its Fourier transform in Fig.~\ref{four0105}. There is a 
QPO with a period near 600 s which is strong in the first half of the run but decreases in
amplitude during the latter half. The average profiles of the QPO for the first and second 
halves are included in Fig.~\ref{QPOqui}. There is some power at the first harmonic of the QPO, 
which shows in the departure from sinusoidality of these mean profiles.

\subsubsection{The interaction of DNOs and QPOs}

\begin{figure}
\centerline{\hbox{\psfig{figure=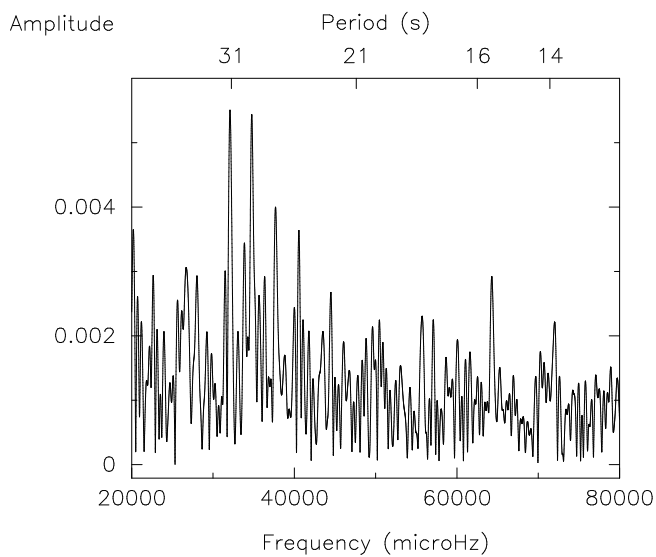,width=8.8cm}}}
  \caption{Fourier spectrum of the first 45 minutes of the light curve on 3 November 1974.}
 \label{four74}
\end{figure}

\begin{figure}
\centerline{\hbox{\psfig{figure=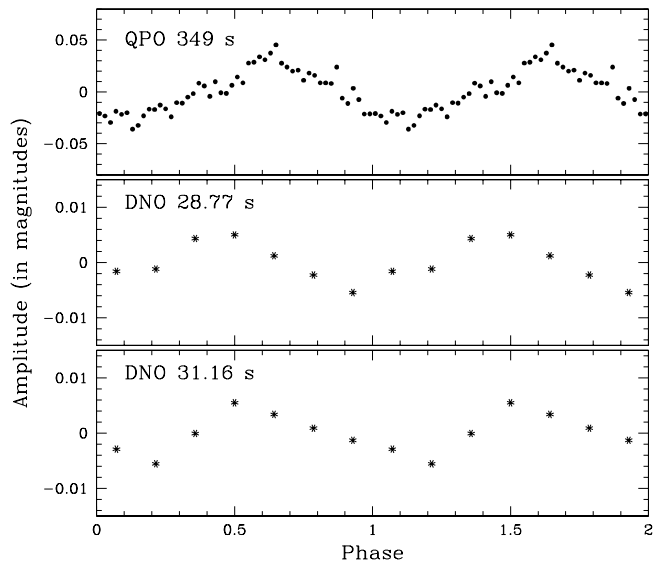,width=8.8cm}}}
  \caption{Averaged profiles of the QPOs and DNOs illustrated in Fig.~\ref{four74}.}
 \label{avprof}
\end{figure}

Of greater interpretative value are the presence of what we will call QPO sidebands.
Fig.~\ref{four74} shows the power spectrum of the first 45 minutes of the light curve obtained on
3 November 1974. The DNO at 28.77 s has a companion at 31.16 s and there is a peak very close
to half the period of the latter, giving physical authenticity to what might otherwise have
been dismissed as a noise spike. The difference frequency of the two DNOs is very close to 
that of the 349 s QPO present in the light curve at that time. The sinusoidal profile of the
28.77 s signal, and the departure from sinusoidality of its companion (confirming the reality
of the harmonic) are shown in Fig.~\ref{avprof}. Clearly the 31.16 s signal is caused by interaction
with the QPO signal -- but is not due to amplitude modulation otherwise there would be two 
sidebands of equal amplitude. The effect is similar to the orbital sideband in intermediate
polars (e.g.~Warner 1986), where the lower frequency signal arises from reprocessing of a 
rotating beam (from the primary) periodically illuminating the secondary or bright spot region,
which makes ``QPO sideband'' an appropriate description. In Paper II we suggest that the
QPO sideband arises from a progradely rotating `wall' in the inner disc.

These signals are only clearly present in the first part of the light curve. From an O--C 
analysis we find that at this time the DNOs show only relatively small jumps in period or phase,
which is what allows the Fourier transform process to detect the signals easily. It is 
possible that in the remainder of this run, and in other similar runs, the QPO sideband
and/or its harmonic may be physically present, but do not stay still long enough to be captured
by our analysis techniques.

In the earlier studies by WB and RW examples were given of the amplitudes of the DNOs
being modulated at the QPO period. We have found several further examples of this, though it is rare
to find both oscillations of sufficient amplitude for this to be readily visible.
There are also examples where the DNO amplitude is unaffected by the QPO modulation.

The examples are too numerous to show in their entirety, but reference to Fig.~\ref{dnobup}
illustrates some of these points. Some of the QPO maxima have large DNO amplitude associated
with them -- in a way that indicates the growth and decay of DNO amplitude over the QPO
maximum. Other QPO maxima have DNOs of low amplitude. DNOs of large amplitude can be seen midway
between the fourth and fifth QPO maxima. DNOs of nearly constant amplitude through a QPO cycle
can be seen in the inset to Fig.~\ref{vwfig1}.

\begin{figure}
\centerline{\hbox{\psfig{figure=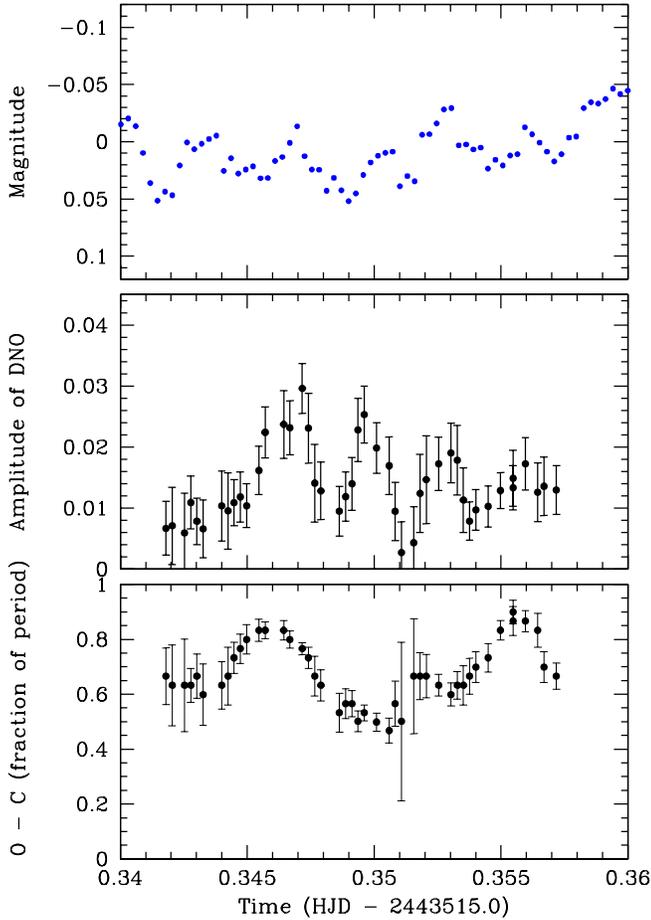,width=8.8cm}}}
  \caption{Analysis of a section of the 6 January 1978 light curve. The upper panel
shows the relative magnitudes of VW Hyi (binned along the horizontal axis by a factor of five, effectively
smoothing over the DNO modulation). The central panel shows the variations in DNO amplitude and the lowest
panel shows the O--C phase variations in which we used 71\% overlap.}
 \label{omcex1}
\end{figure}

\begin{figure}
\centerline{\hbox{\psfig{figure=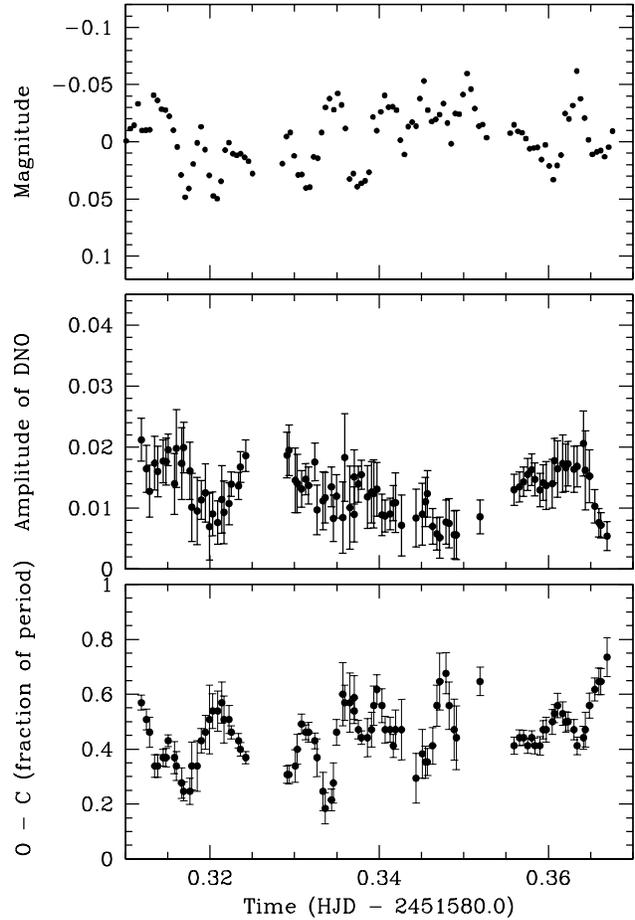,width=8.8cm}}}
  \caption{Analysis of a section of the 5 February 2000 light curve. The upper panel
shows the relative magnitudes of VW Hyi (binned along the horizontal axis by a factor of ten, effectively
smoothing over the DNO modulation). The central panel shows the variations in DNO amplitude and the lowest
panel shows the O--C phase variations in which we used 75\% overlap.}
 \label{omcex2}
\end{figure}

To expand on these examples, and those given in WB and RW, we show detailed analyses in
Figs.~\ref{omcex1} and \ref{omcex2}. Fig.~\ref{omcex1} shows a positive correlation between
DNOs and QPOs, in the sense that the DNOs have maximum amplitude at the peaks of the
QPOs. During this run the DNO phases appear largely independent of the amplitude variations.
In Fig.~\ref{omcex2} the DNO amplitudes appear relatively uncorrelated with the large QPO
modulation, but there is an overall anticorrelation between DNO phase and QPO.
It is certainly noticeable that the time scale of the modulations of DNO phase
is similar to that of the QPO variations.


\section{Concluding Remarks}

This study was stimulated by the lightcurve of VW Hyi at the end of outburst, obtained 
in February 2000 (Fig.~\ref{vwfig1}). At first sight this light curve looks typical of the flickering 
seen in a CV late in outburst.  But closer inspection shows that there is almost no 
flickering present -- the light curve is made up of an orbital modulation plus variable 
amplitude DNOs and QPOs.  The evolution of the DNO and QPO periods in this light 
curve has assisted in selection among the various models of DNOs and QPOs that have 
been proposed.  It is evident that QPOs are more common than realised -- their short 
coherence time results in a broad and noisy signal in the Fourier transform, where (as 
originally pointed out by Patterson et al.~1977), they are easily 
overlooked even though they may be obvious to the eye in the light curve. There is a need 
for an operational definition of QPOs, which can be applied objectively to the light 
curves of CVs.

\section*{Acknowledgments}

We thank the American Association of Variable Star Observers and the Royal
Astronomical Society of New Zealand, specifically and respectively Janet Mattei
and Frank Bateson, for supplying magnitudes of VW Hyi from their archives.
We thank also Darragh O'Donoghue for the use of his EAGLE program.
Many observers have contributed to the VW Hyi archive, in particular, and at our
request, Gerald Handler. BW's research is supported by funds from the 
Research Committee of the University of Cape Town, but not by the National
Research Foundation. Financial support for PAW comes from strategic funds
made available to BW by the University of Cape Town.

\end{document}